\documentclass[aps,twocolumn,superscriptaddress,amsmath,amsfonts]{revtex4-2}

\usepackage{graphicx,float}
\usepackage{braket}
\usepackage{siunitx}
\usepackage{appendix}
\usepackage{amssymb}
\usepackage{mathrsfs}
\usepackage{appendix}
\usepackage{slashed}
\usepackage{physics}
\usepackage{mathtools}
\usepackage{amsmath}
\usepackage{bbold}
\usepackage{xcolor}
\usepackage[mathscr]{euscript}
\usepackage{babel}
\usepackage{tabularx}

\newcommand{\beq}{\begin{equation}}
\newcommand{\eeq}{\end{equation}}

\begin{document}

\title{Quantum nondemolition measurements with optical parametric amplifiers for ultrafast universal quantum information processing}

\newcommand{\SU}{\affiliation{E.\,L.\,Ginzton Laboratory, Stanford University, Stanford, California 94305, USA}}
\newcommand{\NTT}{\affiliation{Physics \& Informatics Laboratories, NTT Research, Inc., Sunnyvale, California 94085, USA}}
\newcommand{\Caltech}{\affiliation{Department of Electrical Engineering, California Institute of Technology, Pasadena, California 91125, USA}}
\newcommand{\MIT}{\affiliation{Research Laboratory of Electronics, MIT, 50 Vassar Street, Cambridge, MA 02139, USA}}

\author{Ryotatsu Yanagimoto} 
\thanks{These authors contributed equally to this work.\\Email: ryotatsu@stanford.edu, rnehra@caltech.edu}
\SU{}
\author{Rajveer Nehra} 
\thanks{These authors contributed equally to this work.\\Email: ryotatsu@stanford.edu, rnehra@caltech.edu}
\Caltech

\author{Ryan Hamerly}
\MIT{}
\NTT{}
\author{Edwin Ng} 
\SU{}
\NTT{}
\author{Alireza Marandi} 
\Caltech
\author{Hideo Mabuchi} 
\SU{}

\begin{abstract}
Realization of a room-temperature ultra-fast photon-number-resolving (PNR) quantum nondemolition (QND) measurement would have significant implications for photonic quantum information processing (QIP), enabling, e.g., deterministic quantum computation in discrete-variable architectures, but the requirement for strong coupling has hampered the development of scalable implementations. In this work, we propose and analyze a nonlinear-optical route to PNR QND using quadratic (i.e., $\chi^{(2)}$) nonlinear interactions. We show that the coherent pump field driving a phase-mismatched optical parametric amplifier (OPA) experiences displacements conditioned on the number of signal Bogoliubov excitations. A measurement of the pump displacement thus provides a QND measurement of the signal Bogoliubov excitations, projecting the signal mode to a squeezed photon-number state. We then show how our nonlinear OPA dynamics can be utilized for \textit{deterministically} generating Gottesman-Kitaev-Preskill states only with additional Gaussian resources, offering an all-optical route for fault-tolerant QIP in continuous-variable systems. Finally, we place these QND schemes into a more traditional context by highlighting analogies between the phase-mismatched optical parametric oscillator and multilevel atom-cavity QED systems, by showing how continuous monitoring of the outcoupled pump quadrature induces conditional localization of the intracavity signal mode onto squeezed photon-number states. Our analysis suggests that our proposal may be viable in near-term $\chi^{(2)}$ nonlinear nanophotonics, highlighting the rich potential of OPA as a universal tool for ultrafast non-Gaussian quantum state engineering and quantum computation.
\end{abstract}

\maketitle 
\section{Introduction}
Quantum information science and engineering offer great potential for revolutionizing many fields such as computation~\cite{Nielsen2000}, communication~\cite{Gisin2007}, and metrology~\cite{LIGO2011}. Among various physical systems that have been experimented with to encode and process quantum information, photonics offers significant advantages in room-temperature scalability and ultra-fast operations~\cite{Obrien2009}. Optical photons can span terahertz bandwidths and propagate over long distances with little decoherence, making them an ideal carrier of quantum information. In photonic quantum computation, information can be encoded and processed in both discrete-variable (DV)~\cite{Obrien2007,Nielsen2004} and continuous-variable (CV)~\cite{Braunstein2005,Menicucci2006} architectures. However, the lack of strong optical nonlinearity has hindered the realization of deterministic two-qubit entangling gates in DV architectures~\cite{Rudolph2017,Slussarenko2019,Li2015}, and non-Gaussian resources such as Gottesman-Kitaev-Preskill (GKP) states~\cite{lloyd1999quantum,Takeda2019,Gottesman2001,Baragiola2019} in CV architectures; both of these are essential for building universal fault-tolerant quantum information processors. While some limitations of weak optical nonlinearity can be circumvented through measurement-based nonlinear operations using photon-number-resolving (PNR) measurement~\cite{Knill2001}, the intrinsically probabilistic nature of these operations and the slow speed of the conventional single-photon detectors (e.g., superconducting nanowires~\cite{Hadfield2009} and superconducting transition-edge sensors~\cite{lita2008counting, nehra2019state}) with complex cryogenic systems severely limit the scalability and computation clock rates in these architectures~\cite{Rudolph2017,Slussarenko2019,Li2015}.

In this context, a realization of ultrafast, room-temperature PNR QND measurement~\cite{Imoto1985, Milburn1983, He2011,Balybin2022} has significant implications in both DV and CV systems. In a PNR QND measurement, information about the number of photons is encoded in an auxiliary probe, and backaction is limited to (partial) projection onto a corresponding photon-number eigenstate~\cite{Grangier1998, Roch1992}. Such an ultrafast QND measurement not only can replace the conventional superconducting PNR detectors, but also can directly realize a deterministic two-qubit entangling gate, which enables deterministic DV optical quantum computation~\cite{Nemoto2004, Nemoto2005,Venkataraman2013}. Additionally, the QND nature of the measurement offers unique opportunities for quantum engineering~\cite{Negretti2007, Geremia2006, Yanagisawa2006}, communication~\cite{Levenson1993}, and metrology~\cite{Kuzmich1998, Shah2010}. To realize a PNR QND measurement, it is typically necessary to engineer a resolvable single-photon energy shift, effectively leading to a strong coupling requirement $g/\kappa>1$ (for coherent coupling rate $g$ and decoherence rate $\kappa$). Since the pioneering works in atom-cavity quantum electrodynamics (QED)~\cite{Nogues1999, Brune1996, Gleyzes2007, Thompson1992, Kimble1998}, strong coupling has been demonstrated in various physical systems~\cite{Englund2007, Wallraff2004, Chu2017}. However, concomitant implementations of the QND measurement in a scalable, high-bandwidth, and room-temperature platform have yet to be developed.

In this work, we propose and analyze a nonlinear-optical route to PNR QND measurements and all-optical quantum state engineering for GKP states using a quadratic optical parametric amplifier (OPA). Compared to the existing PNR QND measurement proposals~\cite{Imoto1985, Milburn1983, He2011,Balybin2022} and GKP-state generation schemes~\cite{Fukui2022,Pirandola2004} using cubic nonlinearities, our proposal with OPA utilizes much stronger quadratic nonlinearity~\cite{Boyd2008}, offering a more experimentally viable route. Recently, $g/\kappa\sim 0.01$ has been demonstrated with a quadratic nonlinear nanophotonic resonator~\cite{Zhao2022,Lu2020}, and even $g/\kappa\sim10$ may be envisioned with ultrafast pulses~\cite{Yanagimoto2022_temporal}. 

In the following, we first show that the pump field of a phase-mismatched OPA experiences conditional displacements depending on the number of signal Bogoliubov excitations $\hat{N}_a$, while $\hat{N}_a$ is approximately preserved under the OPA dynamics. As a result, measuring the pump displacement allows one to perform a PNR QND measurement of $\hat{N}_a$. Next, we show that the nonlinear OPA dynamics can be utilized to perform a modulo quadrature QND measurement~\cite{Weigand2020,Campagne-Ibarcq2020} of the \emph{pump} mode, with which we show a \textit{near-deterministic} generation of the GKP states in the pump mode with only additional Gaussian resources, showing a nonlinear-optical route to universal fault-tolerant CVQIP~\cite{Baragiola2019}. Finally, we bridge the physics of these QND schemes to a more traditional context by establishing analogies between a phase-mismatched optical parametric oscillator (OPO) and multilevel atom-cavity QED systems. We observe conditional localization of the intracavity state to the squeezed Fock state ladder, which in experiments can be inferred from the pump homodyne record without monitoring the signal photon loss at all, using a quantum filter~\cite{Wiseman2009}.

\section{PNR QND Measurements with phase-mismatched OPA}
\begin{figure*}[bth]
    \includegraphics[width=1.0\textwidth]{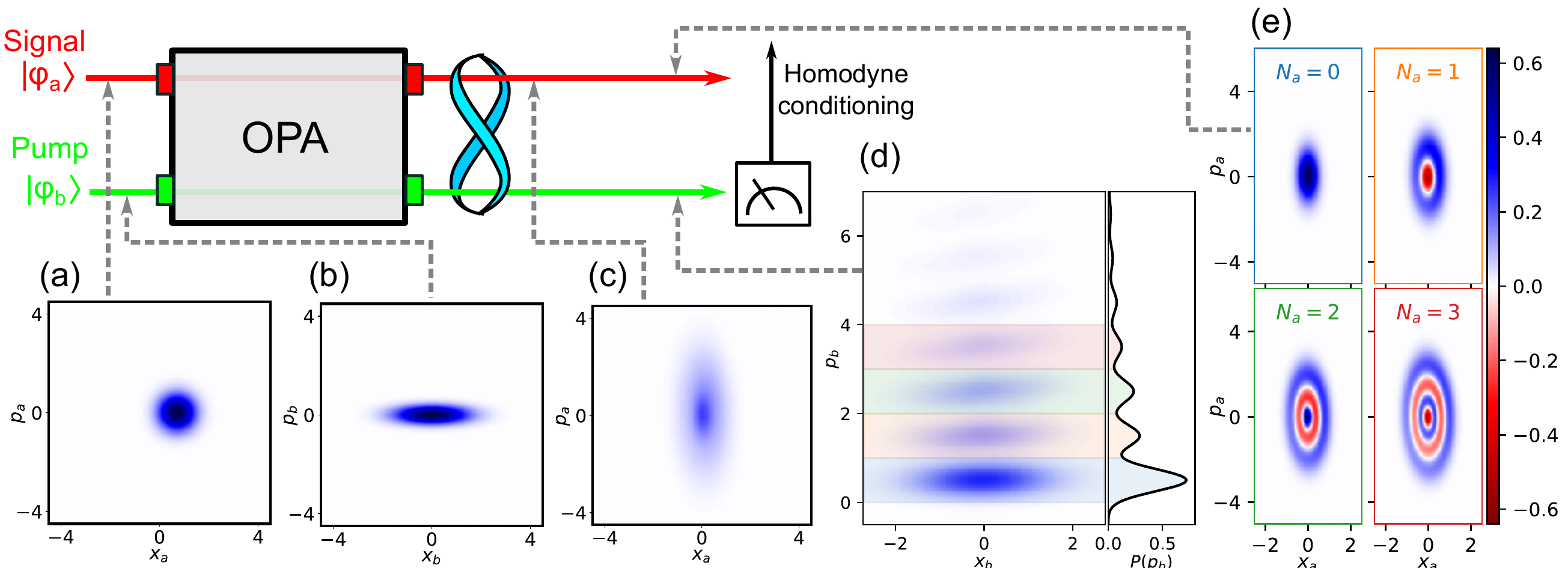}
    \caption{Schematics for our PNR QND measurement scheme using nonlinear quantum behavior of an OPA, where phase-space representation (i.e., Wigner functions) of the system state at each step of the protocol is shown using numerical data. For the numerical simulation, we consider an initial coherent signal state $\ket{\varphi_a(0)}=\ket{\alpha=0.7}$ (shown in (a)), and we assume a $p$-squeezed vacuum state with width $w=1/4$ as an initial pump state (shown in (b)). The signal and pump states interact through a phase-mismatched OPA, whose dynamics induce conditional $p$-displacements to the pump field depending on the number of signal Bogoliubov excitations $\hat{N}_a$. Concurrently, the OPA dynamics also cause conditional rotations on the signal Bogoliubov excitation depending on $\hat{x}_b$, leading to the phase-spread of the final unconditional signal state (shown in (c)). A complete $p$-homodyne measurement on the final pump state acts as a QND measurement of $\hat{N}_a$ and projects the signal mode on a squeezed photon-number state, which is an eigenstate of $\hat{N}_a$. (d) The final pump state shown with the $p$-quadrature distribution $P(p_b)$. (e) Ensemble-averaged signal states conditioned on the outcome of the homodyne measurement within an interval $\tilde{g}t(N_a-1)\leq \hat{p}_b\leq \tilde{g}t(N_a+1)$. We use the system parameters of $\Delta/g=150$ and $\tilde{g}/g=1$, and the total interaction time of $gt=1$.}
    \label{fig:schematics}
\end{figure*}

We consider a phase-mismatched single-mode quadratic (i.e., $\chi^{(2)}$) nonlinear Hamiltonian
\begin{align}
    \label{eq:lab-frame}
    \hat{H}=g(\hat{a}^{\dagger 2}\hat{b}+\hat{a}^2\hat{b}^\dagger)+\delta\hat{a}^\dagger\hat{a},
\end{align}
where $\hat{a}$ and $\hat{b}$ represent annihilation operators for the signal (i.e., fundamental harmonic) and the pump (i.e., second harmonic) modes, respectively, and $g>0$ is the nonlinear coupling strength. We assume a non-negative phase mismatch between signal and pump $\delta\geq 0$ without loss of generality. It is worth noting that various photonic systems can be described by \eqref{eq:lab-frame}, including high-Q microring resonators~\cite{Lu2020}, photonic-crystal cavities~\cite{Medina-Vazquez2022,Wang2020}, temporally trapped ultrashort pulses~\cite{Yanagimoto2022_temporal}, and superconducting microwave circuits~\cite{Krantz2019}, and our results do not rely on a specific physical realization.

To treat the pump coherent amplitude (which may be large in many practical scenarios) in a parametrized way, we transform to a displaced frame given by a unitary $\hat{D}_b(\beta)=\exp(\beta\hat{b}^\dagger-\beta^*\hat{b})$, where the mean field of the pump mode is ``factored out'' as
\begin{align}
    \label{eq:initial}
    \ket{\psi(t)}=\hat{D}_b(\beta)\ket{\varphi(t)},
\end{align}
where $\ket{\psi(t)}$ and $\ket{\varphi(t)}$ are the system states in the lab frame and the displaced frame, respectively. We assume $\beta>0$ without loss of generality. Physically, $\ket{\varphi(t)}$ accounts for quantum fluctuations around the mean field, whose dynamics follow $\mathrm{i}\partial_t\ket{\varphi(t)}=\hat{H}_\text{D}\ket{\varphi(t)}$, where the Hamiltonian
\begin{align}
 \label{eq:displaced-hamiltonian}
    \hat{H}_\text{D}=\hat{D}_b^\dagger(\beta)\hat{H}\hat{D}_b(\beta)=\hat{H}_\text{NL}+\hat{H}_\text{Q}
\end{align}
is composed of a cubic nonlinear term and a quadratic term
\begin{align}
    &\hat{H}_\text{NL}=g(\hat{a}^{\dagger 2}\hat{b}+\hat{a}^2\hat{b}^\dagger),&\hat{H}_\text{Q}=\delta\hat{a}^\dagger\hat{a}+\frac{r}{2}\left(\hat{a}^{\dagger 2}+\hat{a}^2\right)
\end{align}
with $r=2g\beta$. From here on, we assume we are in the displaced frame unless specified.

An OPA is realized for an initial state $\ket{\varphi(0)}=\ket{\varphi_a(0)}\ket{\varphi_b(0)}$ with $\ket{\varphi_b(0)}=\ket{0}$, whose pump state is a coherent state with displacement $\beta$ in the lab frame. A conventional approach to analyze an OPA is to use an undepleted pump approximation, where the pump state remains invariant throughout the dynamics. As shown in Ref.~\cite{Yanagimoto2022-non-gaussian}, this approximation is equivalent to ignoring $\hat{H}_\text{NL}$ in $\hat{H}_\text{D}$, leading to single-mode squeezing of the signal state, which is the expected behavior of an OPA in the regime of Gaussian quantum optics~\cite{Quesada2022}. 

Under stronger nonlinearity where the undepleted pump approximation breaks down, the contribution of the nonlinear term $\hat{H}_\text{NL}$ induces non-Gaussian quantum features, e.g., signal-pump entanglement~\cite{Yanagimoto2022-non-gaussian, Kinsler1993, Xing2022}, for which we critically lack a qualitative physical description. In the following, as a main result of this work, we show a concise description of the nonlinear quantum behavior of phase-mismatched OPA as a QND measurement of signal photons in the squeezed photon-number basis. Our analysis adopts the Hamiltonian transformation recently introduced in Ref.~\cite{Qin2022}.

Assuming a relatively large phase-mismatch $\delta>r$, we can rewrite $\hat{H}_\text{Q}$ as
\begin{align}
    \hat{H}_\text{Q}=\delta\hat{a}^\dagger\hat{a}+\frac{r}{2}\left(\hat{a}^{\dagger 2}+\hat{a}^2\right)=\Delta\hat{A}^\dagger\hat{A}+\mathrm{const},
\end{align}
where $\hat{A}=\hat{a}\cosh u\,+\hat{a}^\dagger\sinh u\,$ corresponds to the annihilation operator for Bogoliubov excitations with $\Delta=\sqrt{\delta^2-r^2}$ and $u=1/2\tanh^{-1}(r/\delta)$. Intuitively, we can interpret $\hat{A}$ as an annihilation operator of a photon excitation in a squeezed photon-number basis. The nonlinear Hamiltonian can then be rewritten in terms of the Bogoliubov operators as
\begin{align}
    \begin{split}
    \hat{H}_\text{NL}&=g\left\{\cosh^2u\,\hat{A}^{\dagger 2}+\sinh^2u\,\hat{A}^2\right.\\&\quad\quad\left.-\sinh 2u\,\left(\hat{A}^\dagger\hat{A}+\frac{1}{2}\right)\right\}\hat{b}+\mathrm{H.c.}
    \end{split}
\end{align}
For the rest of the work, we assume that the magnitude of $\hat{H}_\text{Q}$ dominates over $\hat{H}_\text{NL}$, i.e., $ge^{2u}\ll\Delta$, which can always be achieved by appropriately choosing  $\delta$ and $r$ (i.e., $\beta$). Under these conditions, the contributions from the rapidly rotating terms containing $\hat{A}^2$ and $\hat{A}^{\dagger2}$ average out, allowing us to perform a rotating-wave approximation~\cite{Gu2015,Qin2022,Santamore2004}. We thus have
\begin{align}
    \label{eq:qnd-hamiltonian}
    \hat{H}_\text{D}\approx-2\tilde{g}\left(\hat{N}_a+\frac{1}{2}\right)\hat{x}_b+\Delta\hat{N}_a+\mathrm{const,}
\end{align}
where $\hat{N}_a=\hat{A}^\dagger\hat{A}$, $\hat{x}_b=(\hat{b}+\hat{b}^\dagger)/2$, and $\tilde{g}=g\sinh 2u$. 

In the Heisenberg picture, we analytically solve the operator dynamics under \eqref{eq:qnd-hamiltonian} as 
\begin{align}
    \label{eq:operator}
    &\hat{N}_a(t)\approx\hat{N}_a(0), &\hat{p}_\text{b}(t)\approx\tilde{g}t\left(\hat{N}_a(0)+\frac{1}{2}\right)+\hat{p}_\text{b}(0),
\end{align}
where $\hat{p}_b=(\hat{b}-\hat{b}^\dagger)/2\mathrm{i}$ is the $p$-quadrature operator of the pump mode. From \eqref{eq:operator}, we note that the pump mode $\hat{p}_b$ experiences a  displacement conditioned on the value of $\hat{N}_a$, leading to a specific signal-pump entanglement structure. Additionally, $[\hat{H}_\text{D},\hat{N}_a]\approx 0$ ensures that the value of $\hat{N}_a$ is not disturbed during the system evolution. As a result, homodyne measurement of $\hat{p}_b$ allows us to infer $\hat{N}_a$ \emph{without} performing a destructive measurement on the signal mode, thereby realizing a QND measurement of $\hat{N}_a$. Depending on the measurement result of $\hat{p}_b$, the signal state is projected onto an eigenstate of $\hat{N}_a$ with eigenvalue $N_a$, i.e., a squeezed photon-number state $\ket{N_a}=\frac{1}{\sqrt{N_a!}}\hat{A}^{\dagger{N_a}}\ket{0}$. This situation is summarized in Fig.~\ref{fig:schematics}. 

The performance of our PNR QND measurement depends on the measurement accuracy of $\hat{p}_b$, which is limited by the quadrature fluctuations of the probe pump state. Intuitively, the conditional displacement $d=\tilde{g}t$ needs to be sufficiently large compared to the width of the $p$-quadrature fluctuations $w=\sqrt{\langle\varphi_b\vert\hat{p}_b^2\vert\varphi_b\rangle-\langle\varphi_b\vert\hat{p}_b\vert\varphi_b\rangle^2}$ to infer the value of $\hat{N}_a$ with high confidence. In Fig.~\ref{fig:schematics}, we show the result of a full-quantum simulation of nonlinear OPA dynamics with an initial squeezed-vacuum pump state with $w=1/4$. The final pump state exhibits multiple Gaussian peaks in the phase-space separated by the distance $d$, each of which corresponds to a different number of signal Bogoliubov excitations $\hat{N}_a$. Because we have $d/w=4~(d=1,w=1/4)$ for the parameters used for the figure, conditioning on the measurement result of $\hat{p}_b$ projects the signal state to a squeezed photon-number state with fidelity that can exceed $90\%$ with the assumed system parameters.

\begin{figure}[bt]
\begin{center}
    \includegraphics[width=0.48\textwidth]{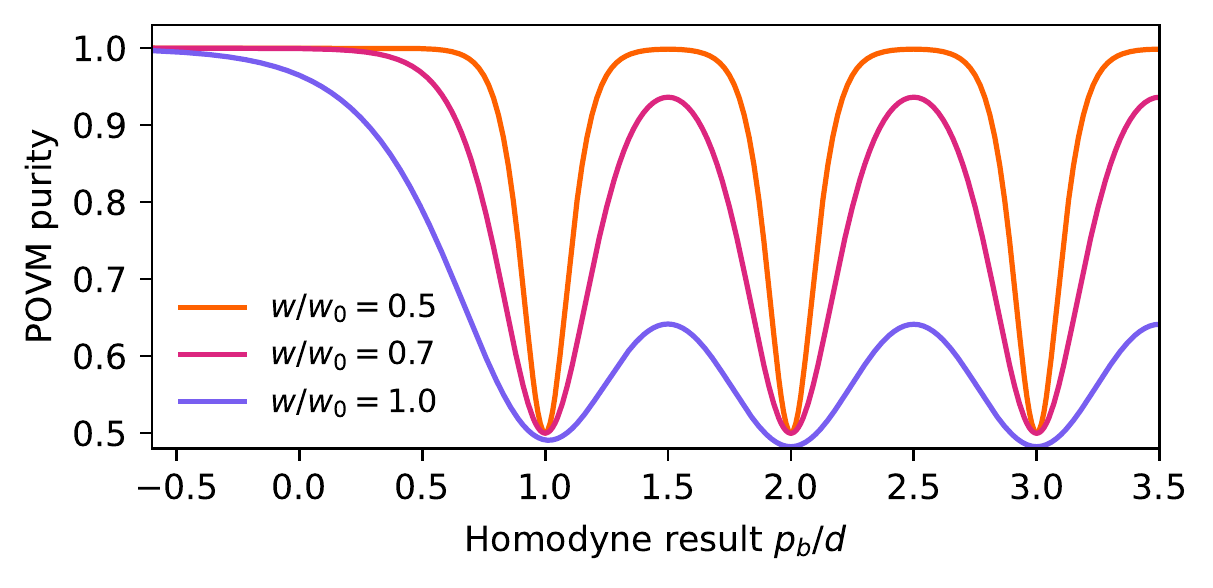}
    \caption{Purity of the POVM for our QND measurement protocol $\mathrm{Pur}(\hat{F}(p_b))=\mathrm{Tr}(\hat{F}^2(p_b))/\mathrm{tr}(\hat{F}(p_b))^2$~\cite{Nehra2020} as a function of the pump homodyne outcome $p_b$. We consider a Gaussian probe pump state $\ket{\varphi_b(0)}$ with various width $w$, where $w$ below the vacuum level $w_0=1/2$ indicates that $\ket{\varphi_b(0)}$ is a squeezed vacuum. We use $d=\tilde{g}t=1.0$ as the amount of conditional displacement.}
    \label{fig:purity}
    \end{center}
\end{figure}

To establish more quantitative connections between the performance of the measurement and the squeezing of the probe-pump quadrature fluctuations, we provide the expressions for the Kraus operators of our QND measurement protocol. From \eqref{eq:qnd-hamiltonian}, the Kraus operators can be expressed as
\begin{align}
\label{eq:povm}
    \hat{M}(p_b)=\sum^\infty_{N_a=0}\mathcal{C}_{N_a}(p_b)\ket{N_a}\bra{N_a}
\end{align}
with $\mathcal{C}_{N_a}(p_b)=e^{-\mathrm{i}\Delta N_at}\left\langle p_b-d\left(N_a+\frac{1}{2}\right)\middle\vert \varphi_b\right\rangle$ being the complex probability amplitudes for the measurement outcomes, where $\ket{p_b}$ is an eigenstate of $\hat{p}_b$ with an eigenvalue $p_b$ (see Appendix~\ref{sec:kraus} for the full derivations). The Kraus operators are related to a positive operator-valued measure (POVM) with elements $\hat{F}(p_b)=\hat{M}^\dagger(p_b)\hat{M}(p_b)$. Physically, the outcome of a complete pump homodyne measurement $p_b$ follows a probability distribution $P(p_b)=\langle\varphi_a\vert\hat{F}(p_b)\vert\varphi_a\rangle$, and conditioned on the outcome $p_b$, the post-measurement signal state becomes $\ket{\varphi_a'}= \hat{M}(p_b)\ket{\varphi_a(0)}$ up to normalization. In Fig.~\ref{fig:purity}, we show the purity of the POVM as a function of the homodyne measurement outcome $p_b$, where we assume squeezed vacuum states with width $w$ as the initial pump state. As can be seen from the figure, use of a pump probe state with smaller $w$ improves the purity of the POVM for a given $d$, projecting the signal to a squeezed photon-number state with a higher fidelity. From an experimental perspective, squeezing the pump quadrature allows us to implement a PNR QND measurement with a shorter nonlinear interaction time, and hence potentially lower propagation loss.

In contrast to the phase-insensitive photon-number tomography attainable by conventional PNR QND measurements~\cite{Imoto1985, Milburn1983, He2011}, our scheme can perform PNR QND measurement in an arbitrary squeezed photon-number basis, enabling phase-sensitive squeeze tomography~\cite{Castanos2004,Ibort2009}, from which we can obtain phase information about the state under tomographic reconstruction. Here, introducing a complex phase to the pump displacement $\beta$ changes the rotation angle of the basis, while the ratio $r/\delta$ determines the squeezing factor. The measurement basis gets more squeezed for $r/\delta\rightarrow1$, where we can have a larger enhancement factor of nonlinear coupling $\tilde{g}/g$. In the other limit of $r/\delta\rightarrow0$, the measurement basis converges to the (non-squeezed) photon-number state basis, which comes with a cost of vanishing effective nonlinear coupling $\tilde{g}/g\rightarrow 0$. It is worth mentioning that additional Gaussian operations can enable flexible control over the measurement basis without compromising the nonlinear coupling. For this purpose, we can apply a pair of opposite squeezing operations $\hat{S}_a$ and $\hat{S}_a^\dagger$ to the signal state before and after evolving under $\hat{H}_\text{D}$, respectively, which transforms the measurement basis so that $\hat{N}_\text{eff}=\hat{A}^\dagger_\text{eff}\hat{A}_\text{eff}$ is measured with $\hat{A}_\text{eff}=\hat{S}^\dagger_a\hat{A}\hat{S}_a$~\cite{Yanagimoto2020}. By choosing $\hat{S}_a$ such that $\hat{A}_\text{eff}=\hat{a}$, we realize a QND measurement of the normal photon number $\hat{n}_a=\hat{a}^\dagger\hat{a}$ without resorting to the limit of $r/\delta\rightarrow0$.

\section{Quantum state engineering for Gottesman-Kitaev-Preskill States}

\begin{figure*}
    \centering
    \includegraphics[width=1.0\textwidth]{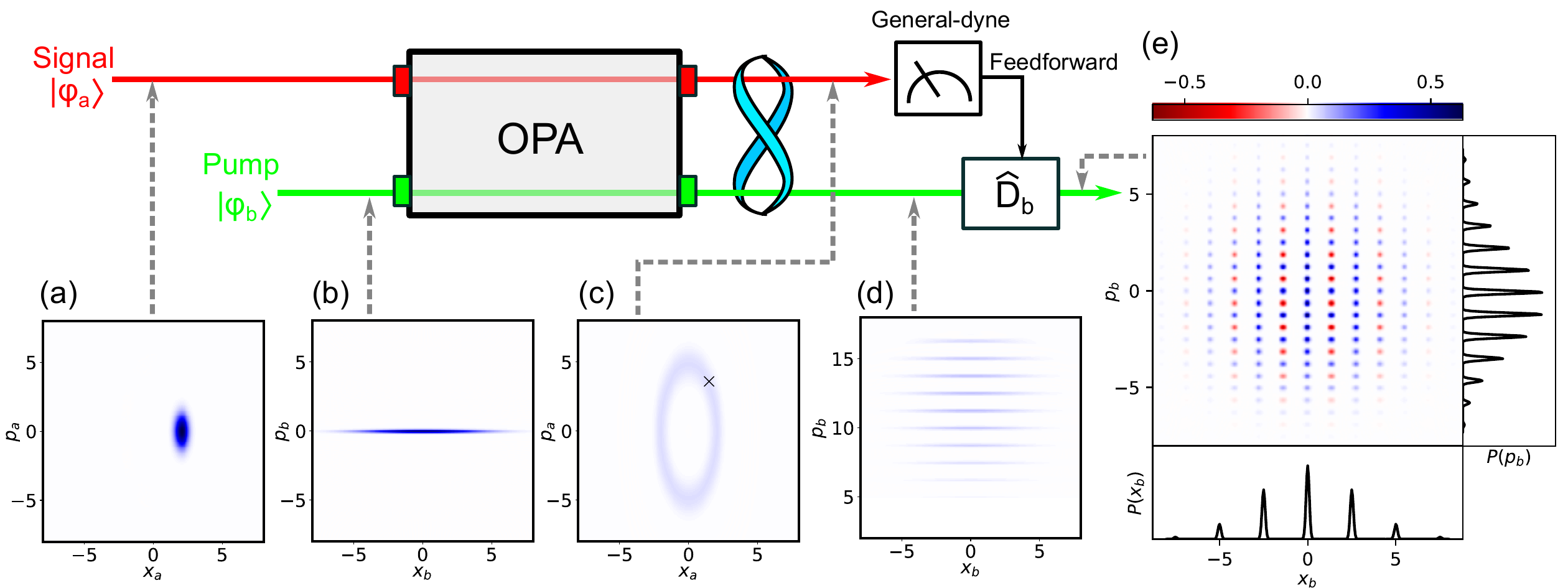}
    \caption{Generation of a symmetric GKP state with $\SI{15}{dB}$ of squeezing using nonlinear quantum dynamics of an OPA. The initial signal state is $\ket{A_0}$ (shown in (a)), and the pump state, which is instantiated as a $\SI{15}{dB}$ of $p$-squeezed vacuum (shown in (b)), interact through a phase-mismatched quadratic OPA Hamiltonian. A complete general-dyne measurement is performed on the final signal mode (shown in (c)) to measure the phase of the signal Bogoliubov excitation $\phi$. Conditioned on the phase measurement outcome, the final pump state (shown in (d)) collapses to an approximate GKP state up to extra displacements, which can be compensated by feedforward operations. The Wigner function and the marginal quadrature distributions of the resultant approximate GKP state are shown in (e). We assume a signal measurement outcome of $\epsilon=0.1$ and $\phi=\pi/4$, whose centroid is indicated by a black cross in (c). For the simulations, we use the Hamiltonian \eqref{eq:qnd-hamiltonian} with $\tilde{g}/g=1.0$ and $\Delta=100$.}
    \label{fig:gkp}
\end{figure*}

While our focus so far has been on QND measurement of the signal excitations, we now show that one can also perform a QND measurement of the pump field quadratures using the same physics of the nonlinear OPA dynamics. For this, we utilize the operator dynamics under \eqref{eq:qnd-hamiltonian} as
\begin{align}
    &\hat{x}_b(t)=\hat{x}_b(0),&\hat{A}(t)=e^{\mathrm{i}(2\tilde{g}t\hat{x}_b(0)-\Delta t)}\hat{A}(0),
\end{align}
where the the information about $2\tilde{g}t\hat{x}_b-\Delta t$ is encoded in the phase of $\hat{A}$ up to the modulo of $2\pi$. Therefore, measuring the phase of $\hat{A}$, e.g., with a general-dyne measurement~\cite{Serafini2017, Wiseman1995,Armen2002}, indirectly infers the value of $\hat{x}_b$ modulo $\mu=\pi/\tilde{g}t$, which projects the pump mode to $\hat{x}_b=x_\phi~(\mathrm{mod}~\mu)$ for a phase measurement outcome of $\phi$, where we denote $x_\phi=(\phi+\Delta t)/2\tilde{g}t~(\mathrm{mod}~\mu)$. The pump quadrature $\hat{x}_b$ itself remains constant throughout the dynamics due to $[\hat{x}_b,\hat{H}_\text{D}]\approx 0$, which ensures QND nature of the measurement. Such modular quadrature measurements play central roles in contemporary CVQIP, e.g., for deterministic generation, stabilization, and quantum error correction with GKP states~\cite{Gottesman2001,Weigand2020,Campagne-Ibarcq2020}. In the following, we demonstrate a preparation of an approximate GKP state using the nonlinear dynamics of an OPA, where only additional Gaussian resources (i.e., Gaussian initial states, measurements, and feedforward operations) are used. Our proposal for generating GKP states adapts Refs.~\cite{Gottesman2001,Weigand2020}, with crucial technical differences stemming from the nonlinear dynamics of phase-mismatched OPAs.

For the following discussions, we denote a coherent excitation of the Bogoliubov signal mode as $\ket{A}$. Physically, $\ket{A}$ is a displaced squeezed state and is an eigenstate of the operator $\hat{A}$ with eigenvalue $A$. As shown in Fig.~\ref{fig:gkp}, we prepare the initial signal state $\ket{A_0}$ with $A_0>0$ as a ``meter'' state for the phase shift. For the initial pump state, we assume a $p$-squeezed vacuum with width $w$ along the $p$-quadrature. After propagating through a nonlinear OPA for time $t$, we perform a phase measurement on $\hat{A}$ by a complete general-dyne measurement~\cite{Serafini2017, Wiseman1995,Armen2002}, which projects the signal mode on the measurement basis of displaced squeezed states $\{\ket{e^{\mathrm{i}\phi}(A_0+\epsilon)}\}$. Here, the measurement basis is parameterized by the radius $(A_0+\epsilon) \geq0$ and the phase $\phi$. For the preparation of a GKP state, a modulo quadrature measurement with modulus $\mu=\sqrt{2\pi}$ is desired, which sets the interaction time $\tilde{g}t=\sqrt{\pi/2}$.

When the magnitude of the meter state $A_0$ is much larger than the vacuum noise level, the measurement outcome is expected to be exponentially localized around $|\epsilon|\ll A_0$. Assuming this condition is met, the post-measurement pump state approximately becomes
\begin{align}
\ket{\varphi_b'}\approx\hat{D}_b(x_\phi)\hat{D}_b\left(\mathrm{i}\sqrt{\pi/2}\,\lfloor A_0\rfloor^2\right)\ket{\tilde{0}},
\end{align}
which can be transformed to an approximate GKP logical state
\begin{align}
    \ket{\tilde{0}}\propto\sum_{n=-\infty}^\infty e^{-\frac{w^2(n\sqrt{2\pi}+x_\phi)^2}{4}}\hat{D}_b(n\sqrt{2\pi})\ket{\kappa}
\end{align}
via trivial displacement operations (see Appendix~\ref{sec:gkp} for more details). Here, $\lfloor\cdot\rfloor$ is a floor function, and $\ket{\kappa}$ is an $x$-squeezed vacuum with width $\kappa=\sqrt{\langle\hat{x}_b^2\rangle-\langle\hat{x}_b\rangle^2}=1/2\sqrt{\pi}A_0$ along the $x$-quadrature. It is worth mentioning that this GKP generation scheme is nearly deterministic, because an extra displacement $\hat{D}_b(\hat{x}_\phi)$ induced by the probabilistic phase readout $\phi$ can be largely compensated by the trivial feedforward displacement operations. The resultant GKP state becomes symmetric when $w=\kappa$ holds true, corresponding to $A_0=1/2\sqrt{\pi}w$.

In Fig.~\ref{fig:gkp}, we show the results of our numerical simulations showing the generation of a symmetric GKP state with squeezing level of $\SI{15}{dB}$ (beyond the error correction threshold of $\sim\SI{10}{dB}$~\cite{Fukui2018,bourassa2021blueprint}). Because a supply of GKP states at one's disposal enables fault-tolerant universal quantum computation with only additional Gaussian resources~\cite{Baragiola2019}, our result shows that a nonlinear OPA is a sufficient component to realize universal nonlinear-optical QC. Compared to existing nonlinear-optical GKP state generation schemes using cross-phase modulation (XPM)~\cite{Pirandola2004,Fukui2022}, our approach employs a much stronger quadratic nonlinearities, which may offer a more viable prospects to non-Gaussian state engineering in room-temperature.

\section{Nonlinear quantum fluctuations in OPO dynamics}
An important application of parametric interactions is an OPO, which is realized by pumping a quadratic nonlinear resonator with an external drive field. In the absence of signal loss, a phase-matched OPO (i.e., $\delta=0$) has two transient states, i.e., odd and even signal cat states comprising of the quantum superposition of $\pi$-phase-shifted coherent states. The presence of a finite signal loss leads to spontaneous switching of the parity of the cat states, devolving the cat states into incoherent mixtures of the original coherent states~\cite{Teh2020, Onodera2022}, which is reminiscent of the spontaneous quantum jumps observed in a two-level atom-cavity QED system~\cite{Mabuchi1998}. Here, we show that a phase-mismatched OPO exhibits behavior reminiscent of \emph{multilevel} atom-cavity QED, where the signal photon loss induces quantum jumps among the signal states in the squeezed Fock state ladder.
 
We introduce an external pump drive for an OPO given by the Hamiltonian term $\hat{H}_\text{drive}=\mathrm{i}\lambda(\hat{b}^\dagger-\hat{b})$, and the outcoupling pump loss is characterized by the Lindblad operator $\hat{L}_b=\sqrt{\kappa_b}(\hat{b}+\beta)$ (In the lab frame $\hat{L}_b=\sqrt{\kappa_b}\hat{b}$). In the absence of signal loss, the pump operator dynamics follow
\begin{align}
    \mathrm{i}\partial_t\hat{b}=-\tilde{g}\left(\hat{N}_a+\frac{1}{2}\right)-\frac{\mathrm{i}\kappa_b}{2}(\hat{b}+\beta)+\mathrm{i}\lambda,
\end{align}
while $\hat{N}_a$ remains constant. For a choice of $\lambda=\kappa_b\beta/2$, we have stationary states $\ket{N_a}\ket{\beta_{N_a}}$ for $N_a\in\mathbb{Z}^+$, where $\ket{\beta_{N_a}}$ is a coherent pump state with displacement $\beta_{N_a}=2\mathrm{i}\kappa_b^{-1}\tilde{g}(N_a+1/2)$. Since $\beta_{N_a}$ depends on $N_a$, the pump photons leaving the OPO carry out information about $N_a$, which plays the role of a weak continuous QND measurement of $\hat{N}_a$. Thus, by monitoring the outcoupled pump field, the system (pump-signal) state is expected to conditionally collapse to one of the stationary states $\ket{N_a}\ket{\beta_{N_a}}$.

Let us now consider the effects of a finite signal loss. When a signal photon is lost from $\ket{\beta_{N_a}}$, the intracavity signal state experiences a quantum jump as $\ket{N_a}\mapsto\hat{a}\ket{N_a}$, resulting in the signal mode given as 
\begin{align}
    \cosh u\,\sqrt{N_a}\ket{N_a-1}-\sinh u\,\sqrt{N_a+1}\ket{N_a+1}.
\end{align}
This implies that a loss of a signal photon, corresponding to a photon subtraction from a squeezed photon-number state, induces a discrete jump of the Bogoliubov excitation $N_a\mapsto N_a\pm 1$, both in the positive and negative directions. Note that the flow is biased towards the negative direction because of $\cosh u>\sinh u$.

\begin{figure}[tb]
    \includegraphics[width=0.5\textwidth]{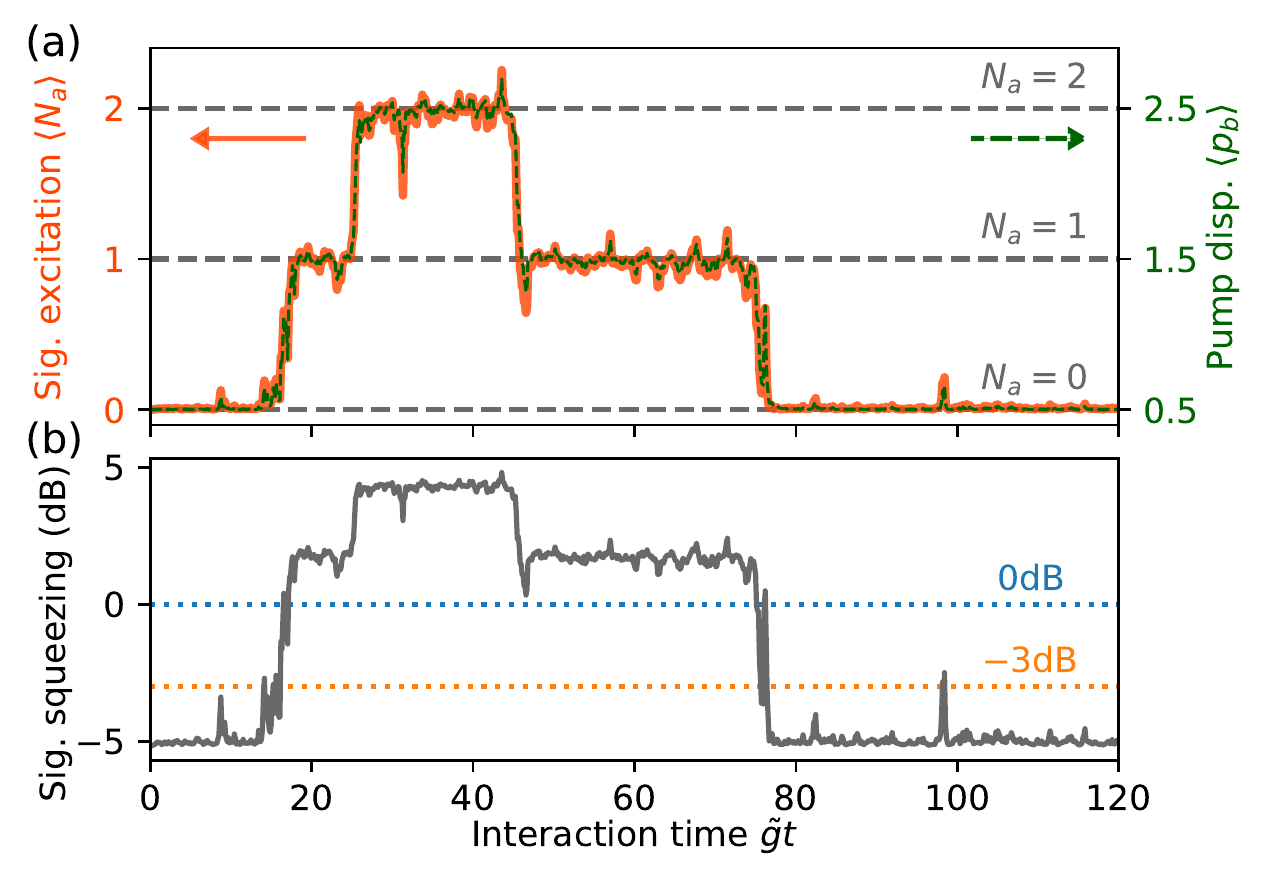}
    \caption{A typical SME quantum trajectory of the OPO dynamics unraveled by a continuous pump $p$-homodyne measurement. (a) Trajectories of $\langle\hat{N}_a\rangle$ (orange solid lines, left axis) and $\langle\hat{p}_b\rangle$ (green dashed lines, right axis) compared to the expected levels of the plateaus $\langle\hat{p}_b\rangle=\mathrm{Im}(\beta_{N_a})$ (grey dashed lines). (b) A trajectory of the signal $x$-quadrature squeezing compared to the quadrature noise levels for a vacuum ($\SI{0}{dB}$, blue dotted line) and the squeezing limit for an OPO steady-state ($\SI{-3}{dB}$, orange dotted line). We use \eqref{eq:qnd-hamiltonian} with system parameters $\Delta/g=100$, $\tilde{g}/g=1.5$, $\kappa_a/g=0.03$, and $\kappa_b/g=3.0$.}
    \label{fig:opo}
\end{figure}

Due to the quantum correlations between $\hat{N}_a$ and $\hat{p}_b$, an occurrence of such a quantum jump can be inferred from the record on the pump homodyne measurement without monitoring the signal loss photons at all. To emulate this situation, we perform numerical simulations of a stochastic master equation (SME)~\cite{Wiseman2010} unraveled by a pump $p$-homodyne measurement, while we do not monitor signal loss photons. As shown in Fig.~\ref{fig:opo}(a), we observe correlated spontaneous jumps in $\langle\hat{N}_a\rangle$ and $\langle\hat{p}_b\rangle$ showing multilevel plateaus corresponding to the production of squeezed photon-number states, which can be inferred solely from the pump homodyne record~\cite{Mabuchi1998,Kerckoff2011}. Such discrete behaviors emerging from a continuous-variable system under the monitoring of only continuous observables are illuminating manifestations of the intrinsic quantum nature of photons. When the system is found in $\ket{N_a=0}\ket{\beta_{N_a=0}}$, the signal state is in a squeezed vacuum, whose squeezing level can conditionally exceed the $\SI{-3}{dB}$ limit of an OPO intracavity steady-state squeezing~\cite{Milburn1981} (see Fig.~\ref{fig:opo}(b)). Note that this phenomena exhibiting strong \emph{signal} squeezing is distinct from the physics unraveled in Ref.~\cite{Qin2022}, where more than $\SI{3}{dB}$ of squeezing is realized in the \emph{pump} mode of an OPO.

\section{Experimental prospects}
We discuss experimental requirements for the implementation of our PNR QND measurement scheme in the single-photon regime. For this purpose, we assume large squeezing factors for all the fields involved in the dynamics, i.e., signal Bogoliubov excitation and probe pump state, to study the potential of squeezing to enhance effective nonlinear coupling. Assuming similar level of loss and squeezing factors for signal and pump, i.e., $\kappa_a\sim\kappa_b$ and $w\sim e^{-u}\ll1$, an experimental requirement for our scheme becomes
\begin{align}
\label{eq:fom}
    \frac{g}{\kappa_\text{a}}\gtrsim w
\end{align}
(see Appendix.~\ref{sec:loss} for full discussions), where the squiggly symbols denote approximate equality (inequality) faithful up to factors of orders of unity. Notice that compared to the normal definition of strong coupling $g/\kappa_a>1$, the requirement \eqref{eq:fom} is reduced by the squeezing of the probe pump mode. For instance, applying $\SI{15}{dB}$ of squeezing on the initial pump can approximately reduce the requirement for $g/\kappa_a$ by the factor of $w^{-1}\sim 5.6$. A promising nonlinear-optical realization of \eqref{eq:lab-frame} is by means of a high-Q microring resonator, where $g/\kappa_a\sim0.01$ has been recently realized in the indium gallium phosphide nanophotonics~\cite{Zhao2022} and the thin-film lithium niobate nanophotonics~\cite{Lu2020}. Moreover, ultrafast pulse operations enabled by advanced dispersion engineering can further enhance the nonlinear coupling by simultaneously leveraging both temporal and spatial field confinements, with which $g/\kappa_a\sim10$ may be possible~\cite{Yanagimoto2022_temporal}. When realized in a single-path manner, such a implementation with ultrashort pulses may enable PNR QND measurements with terahertz through rates. These numbers suggest bright prospects for the potential realization of our proposed scheme on near-term $\chi^{(2)}$ nonlinear nanophotonics.

\section{Conclusion}
In this work, we have proposed and analyzed a scheme for PNR QND measurement and quantum state engineering using the nonlinear quantum behavior of an OPA. We first show that the pump mode driving a phase-mismatched OPA experiences conditional displacements depending on the number of signal Bogoliubov excitations $\hat{N}_a$, enabling one to measure $\hat{N}_a$ nondestructively via a pump homodyne detection. Such PNR QND measurements allow for high-efficiency ultra-fast PNR measurements (replacing the conventional slow superconducting detectors) and a deterministic implementation of photon-photon entangling gate~\cite{Nemoto2004, Nemoto2005}, providing all the necessary elements for deterministic room-temperature DV photonic quantum computation at ultra-fast clock rates. 

We then show that the nonlinear OPA dynamics can be utilized to realize a modular quadrature QND measurement of the pump mode via a signal phase measurement, which naturally provides a way to deterministically generate optical GKP states with additional all-Gaussian resources. Our results unlock many promising opportunities for room-temperature ultra-fast universal quantum computation with GKP states in CV architectures~\cite{Baragiola2019}. It is worth mentioning that our GKP state generation protocol uses Gaussian quadrature measurements, which can be purified using recently demonstrated amplification techniques with high-gain linear OPAs before the inefficient general-dyne measurements, thereby offering a way to generate highly pure GKP states~\cite{Shaked2018,takanashi2020all, nehra2022few}. Finally, extending the discussions to OPO physics, we show that continuous homodyne monitoring of the outcoupled pump field leads to conditional localization of the signal mode on squeezed photon-number states, thereby highlighting a unique opportunity to synthesize and characterize the intracavity nonclassical states in real time.

Our scheme does not rely on materials with cubic nonlinearity, and thus provides a clear path for overcoming the longstanding challenge of the nonlinear-optical PNR QND schemes based on cross-phase modulation (XPM), where the self-phase modulation that inevitably accompanies XPM leads to detrimental phase noise to the probe field~\cite{Imoto1985, Milburn1983, He2011, Balybin2022, Kok2008}. Our work establishes a concise description of the nonlinear-optical parametric interactions beyond the conventional semiclassical picture, thereby showing a practical path toward large-scale, ultrafast, and fault-tolerant universal photonic quantum information processors at room temperatures.

\begin{acknowledgments}
This work has been supported by the National Science Foundation under awards CCF-1918549 and PHY-2011363. R.\,N. and A.\,M. gratefully acknowledge support from NSF grant no. 1846273 and 1918549, AFOSR award FA9550-201-0040, and NASA/JPL. The authors wish to thank NTT Research for their financial and technical support. R.\,Y. is supported by a Stanford Q-FARM Ph.D. Fellowship and the Masason Foundation. 
\end{acknowledgments}

\appendix
\section{Kraus Operators for PNR detection}
\label{sec:kraus}
In this section, we derive the Kraus operators of the PNR QND measurement implemented with the Hamiltonian $\hat{H}_\text{D}$. For the pump $p$-homodyne outcome of $p_b$, post-measurement signal state becomes
\begin{align}
    \ket{\varphi_a'}=\bra{p_b}e^{-\mathrm{i}\hat{H}_\text{D}t}\ket{\varphi_a(0)}\ket{\varphi_b(0)}
\end{align}
up to  normalization, where $\ket{p_b}$ is an engenstate of $\hat{p_b}$ with an eigenvalue $p_b$. For the target signal state $\ket{\varphi_a(0)}=\sum_{N_a=0}^\infty \alpha_{N_a}\ket{N_a}$, we have 
\begin{align}
\begin{split}
\label{eq:post-measurement}
    \ket{\varphi_a'}&=\bra{p_b}\sum_{N_a=0}^\infty \alpha_{N_a}e^{-\mathrm{i}\Delta N_at}\hat{D}_b(\gamma_{N_a})\ket{N_a}\ket{\varphi_b(0)}\\
    &=\sum_{N_a=0}^\infty \alpha_{N_a}C_{N_a}(p_b)\ket{N_a}
    \end{split}
\end{align}
with $\gamma_{N_a}=\mathrm{i}d\left(N_a+\frac{1}{2}\right)$ and
\begin{align}
    C_{N_a}(p_b)=e^{-\mathrm{i}\Delta N_at}\braket{p_b-d(N_a+1/2)}{\varphi_b(0)}.
\end{align}
Here, the equation \eqref{eq:post-measurement} can be summarized as
\begin{align}
    \ket{\varphi_a'}=\hat{M}(p_b)\ket{\varphi_a(0)}
\end{align}
using Kraus operators
\begin{align}
    \hat{M}(p_b)=\sum_{N_a=0}^\infty C_{N_a}(p_b)\ketbra{N_a}.
\end{align}
Assuming a squeezed vacuum with width $w$ along the $p$-quadrature as the pump probe state $\ket{\varphi_b(0)}$, we can analytically write down the complex probability amplitude
\begin{align}
\label{eq:cfunction}
    C_{N_a}(p_b)=\frac{e^{-\mathrm{i}\Delta N_at}e^{-\frac{1}{4w^2}\left(p_b-d\left(N_a+\frac{1}{2}\right)\right)^2}}{(2\pi)^{1/4}w^{1/2}},
\end{align}
which is a Gaussian function centered around $p_b=d(N_a+1/2)$ with width $w$.

The positive valued operator measure (POVM) of the QND measurement protocol $\hat{F}(p_b)$ can be readily obtained from the Kraus operators as
\begin{align}
    \hat{F}(p_b)=\hat{M}^\dagger(p_b)\hat{M}(p_b).
\end{align}
Notice that the POVM fulfills a normalization condition $\int\mathrm{d}p_b~\hat{F}(p_b)=\mathbb{1}_a$.

\section{All-Gaussian generation of GKP states}
\label{sec:gkp}
In this section, we introduce the generation scheme of GKP states by means of a modular quadrature measurement using the nonlinear quantum behavior of an OPA. Our basic approach adapts the protocols using ponderomotive interactions introduced in Ref.~\cite{Gottesman2001,Weigand2020}.

For the following discussions, we denote a coherent excitation of the signal Bogoliubov excitation as $\ket{A}$. Physically, $\ket{A}$ is a displaced squeezed state and is an eigenstate of $\hat{A}$ with an eigenvalue $A$. As an initial system state, we consider
\begin{align}
    \ket{\varphi(0)}=\ket{A_0}\int\mathrm{d}x_b\,\varphi_b(x_b)\ket{x_b},
\end{align}
with $A_0>0$, and $\varphi_b(x_b)$ represents the $x$-quadrature amplitude of the initial pump state. After propagation through a phase-mismatched OPA for time $t$, we measure the phase of the signal mode via a general-dyne measurement~\cite{Serafini2017,Armen2002,Wiseman1995}. This projects the signal state on a measurement basis spanned by states $\{\ket{e^{\mathrm{i}\phi}(A_0+\epsilon)}\}$, where a displaced squeezed state $\ket{e^{\mathrm{i}\phi}(A_0+\epsilon)}$ is parameterized by the radius $A_0+\epsilon\geq 0$ and the phase $\phi$.

For a given measurement outcome of $\epsilon$ and $\phi$, the post-measurement pump state becomes
\begin{align}
    \ket{\varphi_b'}&=\bra{e^{\mathrm{i}\phi}(A_0+\epsilon)}e^{-\mathrm{i}\hat{H}_\text{D}t}\ket{A_0}\ket{\varphi_b(0)}\\
    &=\int\mathrm{d}x_b\,\braket{A_0+\epsilon}{e^{\mathrm{i}(2\tilde{g}tx_b-\Delta t-\phi)}A_0}\varphi_b(x_b)\ket{x_b}\nonumber
\end{align}
up to normalization. As a result, we can write the Kraus operators representing the measurement protocol as
\begin{align}
    \hat{M}(\epsilon,\phi)=\frac{A_0+\epsilon}{\pi}\int\mathrm{d}x_b\,C_{x_b}(\epsilon,\phi)\ketbra{x_b},
\end{align}
where 
\begin{align}
\label{eq:c-amp}
\begin{split}
    C_{x_b}(\epsilon,\phi)=&\exp\left\{-\frac{1}{2}(A_0^2+(A_0+\epsilon)^2\right.\\
    &\quad\quad\left.-2A_0(A_0+\epsilon)e^{\mathrm{i}(2\tilde{g}tx_b-\Delta t-\phi)}\right\}
\end{split}
\end{align}
is a complex amplitude.

When we employ a ``meter'' signal state with an amplitude much greater than the noise level of a vacuum, the outcome of the signal measurement is expected to be exponentially localized around $|\epsilon|\ll A_0$. Assuming that this condition is met, we can approximate \eqref{eq:c-amp} as
\begin{align}
\label{eq:c-amp-approx}
    C_{x_b}(\epsilon,\phi)\approx \sum_{n=-\infty}^\infty&\exp\left\{-2A_0^2(\tilde{g}t)^2(x_b-x_n-x_\phi)^2\right\}\nonumber \\
    &\times \exp\left\{2\mathrm{i}A_0^2\tilde{g}t(x_b-x_n-x_\phi)\right\}
\end{align}
where $x_n=n\mu$ and $x_\phi=(\Delta t+\phi)/2\tilde{g}t~(\mathrm{mod}~\mu)$ with $\mu=\pi/\tilde{g}t$. Notice that \eqref{eq:c-amp-approx} exhibits multiple Gaussian peaks with width $\kappa=1/2\sqrt{\pi}A_0$ separated by an equal distance $\mu$. 

For the generation of a GKP state, we specifically consider a $p$-squeezed pump state
\begin{align}
    \varphi_b(x_b)=\frac{w^{1/2}}{(2\pi)^{1/4}}e^{-\frac{w^2x_b^2}{4}},
\end{align}
whose width along the $p$-quadrature is $w$. Also, we set the interaction time to $\tilde{g}t=\sqrt{\pi/2}$ so that $\mu=\sqrt{2\pi}$. For these parameters, the post-measurement pump state becomes
\begin{widetext}
\begin{align}
\begin{split}
    \ket{\varphi'_b}&\approx \sum_{n=-\infty}^\infty\int\mathrm{d}x_b\,e^{-\frac{w^2x_b^2}{4}}\exp\left\{-2A_0^2(\tilde{g}t)^2(x_b-x_n-x_\phi)^2\right\}\nonumber\exp\left\{2\mathrm{i}A_0^2\tilde{g}t(x_b-x_n-x_\phi)\right\}\ket{x_b}\\
    &\approx \sum_{n=-\infty}^\infty e^{-\frac{w^2(x_n+x_\phi)^2}{4}}\hat{D}_b(x_\phi)\hat{D}_b(x_n)\hat{D}_b(\mathrm{i}\sqrt{\pi/2}A_0^2)\ket{\kappa}\\
    &=\hat{D}_b(x_\phi)\hat{D}_b(\mathrm{i}\sqrt{\pi/2}A_0^2)\sum_{n=-\infty}^\infty e^{-\mathrm{i}\sqrt{2\pi} A_0^2 x_n} e^{-\frac{w^2(x_n+x_\phi)^2}{4}}\hat{D}_b(x_n)\ket{\kappa},
    \end{split}
\end{align}
\end{widetext}
where we have ignored overall normalization constants. Here, $\ket{\kappa}$ is an $x$-squeezed vacuum with width $\kappa$ along the $x$-quadrature. Assuming that $\ket{\kappa}$ is strongly squeezed, we can perform an approximation
\begin{align}
\begin{split}
    &e^{-\mathrm{i}\sqrt{2\pi} A_0^2 x_n}\hat{D}_b(x_n)\ket{\kappa}\\
    &=e^{-2\pi \mathrm{i} n A_0^2}\hat{D}_b(x_n)\ket{\kappa}\\
    &=e^{-2\pi \mathrm{i} n (A_0^2-\lfloor A_0^2\rfloor) }\hat{D}_b(x_n)\ket{\kappa}\\
    &=e^{-\mathrm{i}\sqrt{2\pi} (A_0^2-\lfloor A_0^2\rfloor) x_n}\hat{D}_b(x_n)\ket{\kappa}\\
    &\approx e^{-\mathrm{i}\sqrt{2\pi} (A_0^2-\lfloor A_0^2\rfloor) \hat{x}}\hat{D}_b(x_n)\ket{\kappa}\\
    &=\hat{D}_b\left(-\mathrm{i}\sqrt{\pi/2} (A_0^2-\lfloor A_0^2\rfloor)\right)\hat{D}_b(x_n)\ket{\kappa},
    \end{split}
\end{align}
where $\lfloor\cdot\rfloor$ is a floor function. This allows us to rewrite the post-measurement state as
\begin{align}
\label{eq:disp-gkp}
    \ket{\varphi_b'}\approx \hat{D}_b(x_\phi)\hat{D}_b\left(\mathrm{i}\sqrt{\pi/2}\lfloor A_0^2\rfloor\right)\ket{\tilde{0}}
\end{align}
where
\begin{align}
    \ket{\tilde{0}}\propto \sum_{n=-\infty}^\infty e^{-\frac{w^2(n\sqrt{2\pi}+x_\phi)^2}{4}}\hat{D}_b(n\sqrt{2\pi})\ket{\kappa}
\end{align}
is an approximate GKP logical state. Notice that feedforward displacement operations based on the general-dyne measurement result can transform \eqref{eq:disp-gkp} to an approximate GKP state.

\section{Experimental requirements for the PNR QND measurement}
\label{sec:loss}
In this section, we study the experimental requirements for the implementation of PNR QND measurement scheme in the single-photon regime. In the presence of dissipation, the density matrix for the system state follows the master equation
\begin{align}
\label{eq:master}
    \frac{\mathrm{d}\hat{\rho}}{\mathrm{d}t}=-\mathrm{i}[\hat{H},\hat{\rho}]+\sum_{j\in \{a,b\}}\left(\hat{L}_j^\dagger \hat{\rho}\hat{L}_j-\frac{1}{2}\{\hat{\rho},\hat{L}_j^\dagger\hat{L}_j\}\right),
\end{align}
where $\{\hat{O}_1,\hat{O}_2\}=\hat{O}_1\hat{O}_2+\hat{O}_2\hat{O}_1$ is an anti-commutator. In the main text, we have assumed the dynamical time scale of the phase rotation of the Bogoliubov excitation $\Delta$ dominates over the nonlinear coupling rate, i.e., $\Delta\gg g$. Here, we further assume that $\Delta$ dominates over the time scale of dissipation as well, i.e., $\Delta\gg \kappa_a,\kappa_b$. When this assumption holds, by virtue of the rotating wave approximation, we are justified to ignore contributions from the rapidly rotating terms terms containing $\hat{A}^2$ and $\hat{A}^{\dagger2}$ in \eqref{eq:master}. Specifically, for the terms describing signal loss, we have
\begin{align}
\label{eq:rwa}
    \hat{L}_a^\dagger \hat{\rho}\hat{L}_a-\frac{1}{2}\{\hat{\rho},\hat{L}_a^\dagger\hat{L}_a\}\approx \sum_{j\in\{+,-\}}\hat{L}_j^\dagger \hat{\rho}\hat{L}_j-\frac{1}{2}\{\hat{\rho},\hat{L}_j^\dagger\hat{L}_j\}
\end{align}
with $\hat{L}_+=\sqrt{\kappa_a}\sinh(u)\hat{A}^\dagger$ and $\hat{L}_-=\sqrt{\kappa_a}\cosh(u)\hat{A}^\dagger$. This result indicates that, under a rotating-wave approximation, we can decompose the effect of the original signal Lindblad operator $\hat{L}_a=\sqrt{\kappa_a}\hat{a}$ into that of two Lindblad operators $\hat{L}_+$ and $\hat{L}_-$.

In the following discussions, for concreteness, we consider a squeezed single-photon state $\ket{N_a=1}$ as an initial signal state. For the initial pump state, we assume $p$-squeezed vacuum with width $w$ along the $p$-quadrature. For a successful PNR QND measurement, the probability for a quantum jump to occur in the signal mode should be sufficiently low. In the low-loss limit, the probability for a quantum jump is approximately given as
\begin{align}
    P_\text{jump}&=1-\langle N_a=1\vert e^{-\hat{L}_a^\dagger\hat{L}_at}\vert N_a=1\rangle \nonumber\\
    &\approx 1-\langle N_a=1\vert e^{-(\hat{L}_+^\dagger\hat{L}_++\hat{L}_-^\dagger\hat{L}_-)t}\vert N_a=1\rangle \nonumber\\
    & = 1-\langle N_a=1\vert e^{-\kappa_a (\cosh(2u)\hat{N}_a+\sinh^2(u))t}\vert N_a=1\rangle \nonumber\\
    &= 1-e^{-\kappa_a(3\cosh^2(u)-2)t}
\end{align}
which sets a characteristic timescale for the loss-induced quantum jump as $t_\text{jump}\sim 1/\cosh^2(u)\kappa_a$. 

To be able to measure $\hat{N}_a$ with high confidence, the conditional displacement occurring over the time scale of $t_\text{jump}$ needs to be greater than the characteristic width of the pump state. Now, in the presence of finite but small pump loss, the width of the final pump state along the $p$-quadrature becomes
\begin{align}
\begin{split}
    w'(t)&=\sqrt{w^2e^{-\kappa_b t}+(1-e^{-\kappa_b t})/4}\\
    &\approx \sqrt{w^2+(1/4-w^2)\kappa_b t},
\end{split}
\end{align}
where we have assumed $\kappa_b t\ll1$. As a result, experimental condition for a successful implementation of our scheme becomes $\tilde{g}t_\text{jump}\gtrsim w'(t_\text{jump})$. Here, we use the squiggly symbol to denote approximate equality up to factors with orders of unity.

Here, we assume strong squeezing for all the fields involved, i.e., signal Bogoliubov excitation and the pump state. Also, we assume similar level of loss and squeezing for both signal and pump, i.e., $\kappa_a\sim\kappa_b$ and $w\sim e^{-u}\ll1$. Under these conditions, the order of magnitude of $w'(t_\text{jump})$ is larger than $w$ only by a factor of unity, allowing us to approximate $w'(t_\text{jump})\sim w$. As a result, we obtain a concise expression for the experimental requirement for the our scheme as
\begin{align}
    \frac{g}{\kappa_a}\gtrsim w.
\end{align}

\bibliography{myfile}
\end{document}